\documentclass{ws-p8-50x6-00}
 
\usepackage{graphics}

\renewcommand{\dj}{\hbox{d\hskip-1.1ex{\raise0.640ex\hbox{--}}\skip 0.70ex}}

\begin{document}



\begin{flushright}
IRB-TH-8/99\\
November, 1999
\end{flushright}

\vspace{2.0cm}

\begin{center}
\Large\bf Inclusive decays and lifetimes of doubly charmed baryons  
\vspace*{0.3truecm}
\end{center}

\vspace{1.8cm}

\begin{center}
\large B. Guberina, B. Meli\'{c} and
 H. \v Stefan\v ci\'c\\
{\sl Theoretical Physics Division, Rudjer Bo\v skovi\'c Institute,
P.O.Box 1016, HR-10001 Zagreb, Croatia\\[3pt]
E-mails: {\tt guberina@thphys.irb.hr, 
         \tt melic@thphys.irb.hr,
         \tt shrvoje@thphys.irb.hr}}

\end{center}

\vspace{1.5cm}

\begin{center}
{\bf Abstract}\\[0.3cm]
\parbox{13cm}
{
We extend the analysis of weak decays of heavy hadrons
to the case of doubly charmed baryons, $\Xi_{cc}^{++}$, $\Xi_{cc}^{+}$
and $\Omega_{cc}^{+}$. Doubly charmed baryons are
modeled as a heavy-light system containing a heavy $cc$-diquark
and a light quark. Such a model leads
to preasymptotic effects in semileptonic and nonleptonic
decays which are essentially proportional to the {\it meson} wave
function. Very clear predictions
for semileptonic branching ratios and lifetimes
of doubly charmed baryons are obtained.
}
\end{center}

\vspace{2.0cm}

\begin{center}
{\sl Talk given by H. \v Stefan\v ci\'c at
 the XVII Autumn School "QCD; Perturbative or Nonperturbative ?",
Lisbon, Portugal, 29 Sept - 4 Oct 1999.\\
To appear in the Proceedings}
\end{center}

\thispagestyle{empty}
\vbox{}
\newpage

\setcounter{page}{1}


\title{Inclusive decays and lifetimes of doubly charmed baryons} 

\author{B. Guberina, B. Meli\'{c} and
 H. \v Stefan\v ci\'c\thanks{Talk given by H. \v Stefan\v ci\'c at
 the XVII Autumn School "QCD; Perturbative or Nonperturbative ?", 
Lisbon, Portugal, 29 Sept - 4 Oct 1999.}
}                    

\address{
  Theoretical Physics Division, Rudjer Bo\v{s}kovi\'{c} Institute, \\
   P.O.B. 1016, HR-10001 Zagreb, Croatia}


\maketitle

\abstracts{We extend the analysis of weak decays of heavy hadrons 
to the case of doubly charmed baryons, $\Xi_{cc}^{++}$, $\Xi_{cc}^{+}$
and $\Omega_{cc}^{+}$. Doubly charmed baryons are
modeled as a heavy-light system containing a heavy $cc$-diquark 
and a light quark. Such a model leads 
to preasymptotic effects in semileptonic and nonleptonic
decays which are essentially proportional to the {\it meson} wave
function. Very clear predictions 
for semileptonic branching ratios and lifetimes
of doubly charmed baryons are obtained. }

\section{Introduction}
 
In the last decade, significant progress in weak decays of 
heavy hadrons has been achieved \cite{Bigi}. The qualitative picture of the 
lifetime hierarchy predicted for singly charmed baryons and 
charmed mesons has been found to be in agreement with experiments. 

On the other hand, although the inverse bottom-quark mass appears 
to be a better expansion parameter of the inclusive decay formalism,
it seems that the 
predictions in beauty decays do not quite follow the success from the charmed 
sector, especially in the prediction of absolute lifetimes of 
beauty particles \cite{GMS2,Voloshin}.

The following natural step towards the 
investigation of weak-decay dynamics is therefore 
the consideration of heavy baryons 
containing two charmed quarks. The triplet of doubly charmed baryons 
($\Xi_{cc}^{++}$, $\Xi_{cc}^{+}$ and $\Omega_{cc}^{+}$) 
exhibits significant preasymptotic effects, already found 
to be important in singly charmed decays. 
 
We discuss some special features coming from the doubly heavy nature of 
doubly charmed baryons and give predictions for their lifetimes and 
semileptonic branching ratios (BR) \cite{GMS}.  
Some wrong prefactors in \cite{GMS} for the four-quark contributions 
are corrected and numerical results are reevaluated \cite{Err}.

\section{Preasymptotic effects and the wave function in doubly charmed 
baryon decays}

The decay rate of a doubly charmed baryon is expanded through the OPE technique, 
similarly as for a singly charmed baryon, in the series of the product of 
the short-distance part 
(Wilson coefficient functions denoted by $c_i^f$) 
and the long-distance part (matrix elements 
$\sim \langle H_{cc}|{\cal O}_i | H_{cc} \rangle$): 
\begin{eqnarray}
\label{eq:gen}
 \Gamma (H_{cc} \rightarrow f) &=& \frac{G_{\rm F}^2 m_c^5}{192 \pi^3} |V|^2 
\frac{1}{2 M_{H_{cc}}} \{ c_3^f \langle H_{cc}|\overline{c}
c|H_{cc}\rangle \\ \nonumber 
&+& c_5^f \frac{ \langle H_{cc}|\overline{c} g_s \sigma^{\mu \nu} G_{\mu \nu} c|
H_{cc}\rangle }{m_c^2} \\ \nonumber
 &+& \sum_i c_6^f \frac{ \langle H_{cc}| (\overline{c} 
\Gamma_i q)(\overline{q} \Gamma_i c)|H_{cc}\rangle }{m_c^3} + O(1/m_c^4) \}\,. 
\end{eqnarray}

The matrix elements are specific for a given doubly charmed hadron. 
The leading $O(m_{c}^{5})$ term is given by the HQET
expression:
\begin{equation}
\langle H_{cc}|\overline{c}c|H_{cc}\rangle = 
1 - \frac{1}{2}\frac{\mu_{\pi}^2(H_{cc})}{m_c^2}+
     \frac{1}{2}\frac{\mu_G^2(H_{cc})}{m_c^2} \, ,
\end{equation}
where $\mu_G^2$ parametrizes the matrix element of the chromomagnetic
operator $\overline{c} g_s \sigma^{\mu \nu} G_{\mu \nu} c$ and 
$\mu_{\pi}^2$ is the matrix element of the kinetic energy operator.

The value of the kinetic energy matrix element is obtained using some
phenomenological features of the meson potential \cite{Kis}:
\begin{equation}
\label{eq:mupi}
\mu_{\pi}^{2}=m_{c}^{2}v_{c}^{2}=(\frac{m_{q}^{*} T}
{2 m_{c}^{*2}+m_{c}^{*} m_{q}^{*}}+\frac{T}{2 m_{c}^{*}})m_{c}^{2} \, ,
\end{equation}
where T is the average kinetic energy of a light quark and a heavy diquark.
In all
expressions in the paper, $m$ refers to the current mass that is used 
in the HQET and
OPE expansions, while $m^{*}$ refers to the constituent mass used in
model calculations.

The expression for $\mu_{G}^{2}$ is the following
\begin{equation}
\label{eq:mug}
\mu_{G}^{2}=\frac{2}{3}(M_{ccq}^{*}-M_{ccq})m_{c}-
(\frac{2}{9} g_{S}^{2} \frac{|\phi(0)|^2}{m_{c}^{*}}+
\frac{1}{3} g_{S}^{2} \frac{|\phi(0)|^2}{m_{c}}) \, .
\end{equation}
The first term describes the hyperfine interaction between a light quark
and a heavy diquark, while the second one accounts for the hyperfine
interaction of heavy quarks in a heavy diquark. $\phi(0)$ is the
wave function of the $cc$ pair in a diquark.

The second term in the expression (1) describing the chromomagnetic interaction, 
receives a similar contribution from (4). 

The third term in (1) describes four-quark interactions, specific for a given 
hadron. Four-quark operators produce the effect which is the largest of all
preasymptotic effects and numerically
comparable with the leading "decay" contribution. Therefore, these
effects
introduce the crucial difference in lifetimes and semileptonic
branching ratios between various doubly charmed baryons. We state
their contributions explicitly below. 

In the case of semileptonic decays, the contributions of four-quark
operators appear through the positive Pauli interference \cite{Vol}
and can be expressed using 
\begin{equation}
\label{eq:Voloshin}
\tilde{\Gamma}_{SL} = \frac{G_F^2}{12\pi} m_c^2 
(4 \sqrt{\kappa}-1) \frac{10}{3} |\psi(0)|^2\,. 
\end{equation}
For individual doubly charmed baryons, the contributions are the
following:
\begin{eqnarray}
\label{eq:slVol}
\Gamma_{SL}^{4q}(\Xi_{cc}^{++}) &=& 0 \,,\nonumber\\
\Gamma_{SL}^{4q}(\Xi_{cc}^{+}) &=&   s^2 \tilde{\Gamma}_{SL} \,,\nonumber\\
\Gamma_{SL}^{4q}(\Omega_{cc}^{+}) &=&  c^2 \tilde{\Gamma}_{SL}\,. 
\end{eqnarray} 
Expressions for the contributions of four-quark operators 
become more intricate in the case of nonleptonic decays owing to
the more complex $QCD$ dynamics. In general, there are three types
of processes known as $W$ exchange, negative Pauli interference and
positive Pauli interference, given by the following expressions:
\begin{eqnarray}
\label{eq:four}
\Gamma^{ex} &=& \frac{G_F^2}{2\pi} m_c^2 [ c_{-}^2 + \frac{2}{3}(1 - 
\sqrt{\kappa})(c_{+}^2 - c_{-}^2)]\,6|\psi(0)|^2\,, \nonumber \\
\Gamma^{int}_{-} &=& \frac{G_F^2}{2\pi} m_c^2 [ -\frac{1}{2} c_{+}(2 c_{-}-
c_{+}) \nonumber \\ 
& & -  \frac{1}{6}(1 - \sqrt{\kappa})(5 c_{+}^2 + c_{-}^2-6 c_{+} c_{-})]
\,\frac{10}{3}|\psi(0)|^2 \,,\nonumber \\
\Gamma^{int}_{+} &=& \frac{G_F^2}{2\pi} m_c^2 [ \frac{1}{2} c_{+}(2 c_{-}+
c_{+})  \nonumber \\
& & -  \frac{1}{6}(1 - \sqrt{\kappa})(5 c_{+}^2 + c_{-}^2+6 c_{+} c_{-})]
\,\frac{10}{3}|\psi(0)|^2 \, .
\end{eqnarray}
These effects combine to give the contributions of four-quark 
operators to
the triplet of doubly charmed baryons
\begin{eqnarray}
\label{eq:nltot}
\Gamma_{NL}^{4q}(\Xi_{cc}^{++}) 
& = &\{(c^4+s^4)P_{int}(x) 
 +c^2s^2(1+\tilde{P}_{int}(x))\}\Gamma_{-}^{int} \, , \nonumber \\
\Gamma_{NL}^{4q}(\Xi_{cc}^{+}) 
& = &(c^4 P_{ex}(x)+c^2 s^2) \Gamma^{ex} 
+(s^4 P_{int}(x)+c^2 s^2)\Gamma_{+}^{int} \, ,\nonumber \\
\Gamma_{NL}^{4q}(\Omega_{cc}^{+}) 
& = & (c^4+c^2 s^2 P_{int}(x))\Gamma_{+}^{int} 
 +(c^2 s^2 P_{ex}(x)+s^4)  \Gamma^{ex} \, .
\end{eqnarray}     
Here $s^2$ and $c^2$ stand for $sin^2 \theta_c$ and $cos^2 \theta_c$, 
respectively, and 
$|\psi(0)|$ denotes the baryon wave function. 

In the calculation of the baryon wave function the application of
the nonrelativistic quark model \cite{Ruj} gives
\begin{equation}
\label{eq:psi}
|\psi(0)|^2 = \frac{2}{3} |\psi(0)|^2_D =
\frac{2}{3} \frac{f_D^2 M_D \kappa^{-4/9}}{12}.
\end{equation}
In the calculations, the physical value of $f_{D}$ has been  used
(instead of the static value $F_{D}$), consistent
with the considerations on the "mesonic" nature of doubly heavy baryons
\cite{GMS,ShifBlok}. Since the calculation of this wave function 
relies upon the nonrelativistic
quark model, the scale at which the contribution of four-quark
operators is defined is lowered to the typical hadronic scale
($\mu \sim 0.5 - 1 \, GeV$) by the
process of hybrid renormalization, as quark models are supposed to
work best at these scales.

In the calculations of inclusive decay rates all Cabibbo modes
(including the suppressed ones) were taken into account. $QCD$
corrections \cite{Kim,Nir,Bagan}
were calculated for the case of decay diagrams, while
the masses of the particles in the final states were accounted for
by the inclusion of appropriate mass corrections
\cite{Cormass,Hok}.

\section{Semileptonic inclusive rates and lifetimes - results
and discussions}
 
\begin{figure*}[!t]
\centerline{\resizebox{1.0\textwidth}{!}{\includegraphics{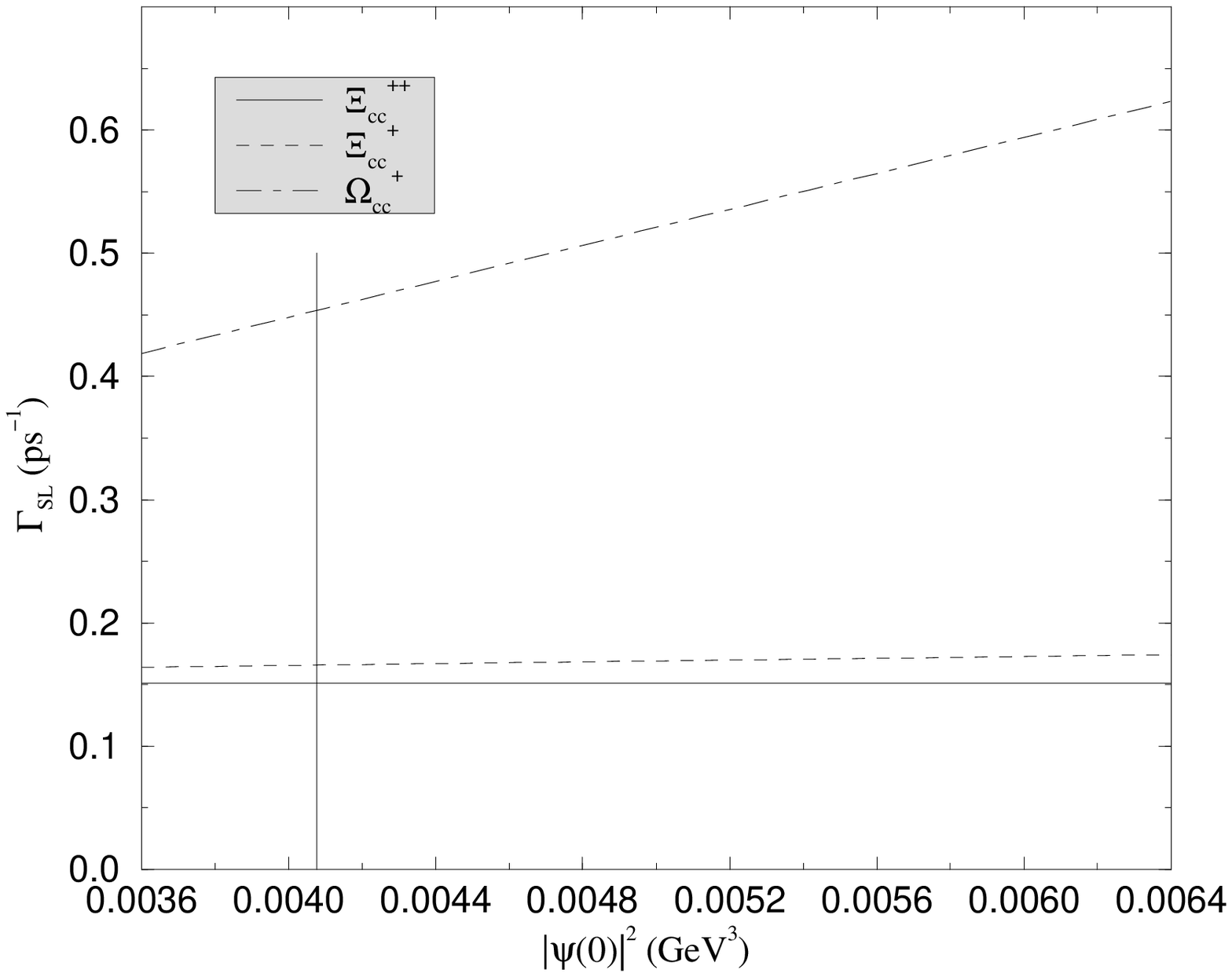}
\includegraphics{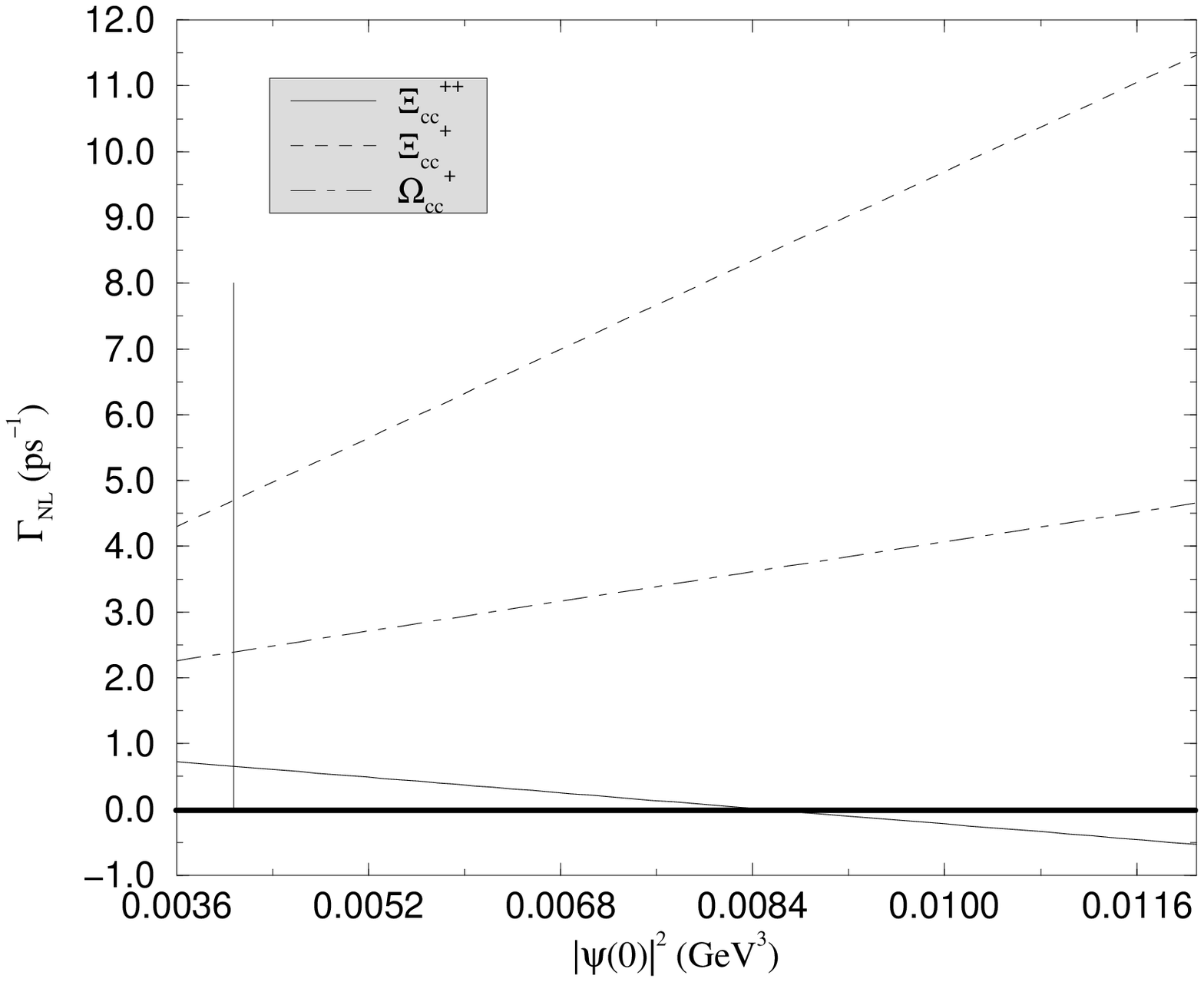}}}
\caption{\label{fig1} Dependence of semileptonic 
and nonleptonic decay widths 
on the value of the wave function squared. The 
 vertical line represents the $\mid \psi(0) \mid^2$
used in calculations, which corresponds to $f_{D}=170 MeV$.}
\end{figure*}

Numerical results are given in Table \ref{tab1}, together with the set of
parameters used in numerical calculations. The complete list of 
parameters can be found in \cite{GMS}. 
\begin{table}
\centering
\begin{tabular}{|c|c|c|c|} \hline \hline
& $  \;\;\;\;\;\;\;\;\;\Large{ \Xi_{cc}^{++}} \;\;\;\;\;\;\;\;\;$
& $\;\;\;\;\;\;\;\;\; \Large{\Xi_{cc}^{+}} \;\;\;\;\;\;\;\;\;$
 & $\;\;\;\;\;\;\;\Large{\Omega_{cc}^{+}} \;\;\;\;\;\;\;$ \\ \hline \hline
\multicolumn{4}{|c|}{Nonleptonic widths in $ps^{-1}$ } \\ \hline
$ \Gamma_{NL}$ & 0.655 & 4.699 & 2.394 \\ \hline \hline
\multicolumn{4}{|c|}{Semileptonic widths in  $ps^{-1}$ }  \\ \hline
$\Gamma_{SL}$ & 0.151 &   0.166
  &  0.454 \\ \hline \hline
\multicolumn{4}{|c|}{Semileptonic branching ratios in \%} \\ \hline
$BR_{SL}$ &  15.8    & 3.3
 & 13.7 \\  \hline \hline
\multicolumn{4}{|c|}{Lifetimes in ps} \\ \hline
$\tau$ &   1.05  &  0.20
&  0.30 \\ \hline \hline
\end{tabular}
\caption{\label{tab1} Predictions for nonleptonic widths, 
semileptonic widths, semileptonic branching
ratios (for one lepton species) and lifetimes of doubly charmed 
baryons for the values of the
parameters $m_{c}=1.35 GeV$, $\mu=1 GeV$, $\Lambda_{QCD}=300 MeV$, $f_{D}=170
MeV$.}
\end{table}
Numerical results show that the dependence on $\mu$ and $\Lambda_{QCD}$
is very weak in the case of $\Xi_{cc}^{++}$, while it is negligible
in the case of $\Xi_{cc}^{+}$ and $\Omega_{cc}^{+}$.
The dependence of semileptonic and nonleptonic decay rates on 
the baryonic
wave function at the origin is shown in Figure \ref{fig1}. 
In the case of $\Omega_{cc}^{+}$, the large contributions of 
four-quark operators make $\Gamma_{SL} (\Omega_{cc}^{+})$ and 
$\Gamma_{NL} (\Omega_{cc}^{+})$ strongly $|\psi(0)|^2$ dependent. 

In the case of nonleptonic total decay rates, our choice of $f_{D}$
in calculation of $\psi(0)$ is numerically confirmed,   
because the
choice of static value $F_{D}$ would cause the
$\Gamma_{NL}(\Xi_{cc}^{++})$ to become negative, what is an obviously 
unphysical result.
One can summarize numerical results in the form of hierarchies of
relevant calculated quantities. 
The relation for semileptonic branching ratios is
\begin{equation}
\label{eq:brsl}
BR_{SL}(\Xi_{cc}^{+}) \ll BR_{SL}(\Omega_{cc}^{+}) <
BR_{SL}(\Xi_{cc}^{++}) \, , 
\end{equation}
while the lifetimes satisfy the following pattern:
\begin{equation}
\label{eq:tau}
\tau(\Xi_{cc}^{+}) \sim \tau(\Omega_{cc}^{+}) \ll
\tau(\Xi_{cc}^{++}) \,.
\end{equation}
The lifetime of $\Xi_{cc}^{++}$ is strongly prolonged by the inclusion 
of the negative Pauli interference effects.

Finally, one can estimate the lifetime of the triply charmed baryon
$\Omega_{ccc}^{++}$ using the analogous procedure applied 
in the preceding consideration.
Since there are no light valence quarks in the structure of $\Omega_{ccc}^{++}$, 
four-quark operators give no contribution, and
the lifetime can be estimated using only the leading term in the $1/m_{c}$
expansion. Such an approach gives \cite{GMS}
\begin{equation}
\tau(\Omega_{ccc}^{++})=0.43 \, ps \,.
\end{equation}

\section{Conclusions}

An interesting hierarchy of semileptonic branching ratios and
lifetimes of doubly charmed baryons has been predicted. The large spread
of results indicates the importance of preasymptotic effects. The
level of agreement of these predictions with experimental results
of future experiments will measure the applicability of the formalism 
to doubly heavy systems and will possibly shed some light on the
validity of some underlying assumptions, such as quark-hadron
duality.

{\em Acknowledgements.} This work was supported by the Ministry of Science and
Technology of the Republic of Croatia under Contract No. 00980102

\end{document}